# Analytical models of operational risk and new results on the correlation problem

Vivien BRUNEL

Head of Risk and Capital Modeling

Société Générale

vivien.brunel@socgen.com

(this version: April 4[th], 2014)

We propose a portfolio approach for operational risk quantification based on a class of analytical models from which we derive new results on the correlation problem. In particular, we show that uniform correlation is a robust assumption for measuring capital charges in these models.

Introduction

The regulatory framework allows banks to compute their capital charge for operational risk under an internal model, which is often based on the Loss Distribution Approach (LDA). In this approach, loss distributions are calibrated at the cell level (a cell is the elementary risk unit per business line and type of risk) and the bank's capital charge is estimated by aggregating cell loss distributions under some dependence assumption (Chernobai *et al.*, 2007).

The Basel Committee (Bank for international Settlements, 2011) provides some guidelines about how banks should appropriately reflect the risk profile in their internal model. However, banks benefit from some flexibility in their modeling choices that may lead to some discrepancies in capital charges for similar risk profiles. The broad range of practices observed among banks results in particular from different distributional or dependence assumptions in the models.

Many studies have focused on the modeling of the tails in the severity distributions (Dutta and Perry, 2007; Moscadelli, 2004) but the bulk of the correlation problem is still unsolved and controversial. There is a strong debate about the choice of the copula function for losses across cells, since the scarcity of the data prevents from solving this issue. The regulator advises banks to determine sound correlations and to retain conservative assumptions. Some institutions have selected the simplest option and use equal correlations between cell losses. This assumption is of course questionable and may embed some model risk, but regulators, as well as practitioners, have great difficulties in asserting arguments about realistic and conservative correlation levels. Some authors believe that correlations between cell losses are as low as 4% (Frachot *et al.*, 2004).

Most of the knowledge we have about operational risk quantification comes from complex models and heavy Monte-Carlo simulations, and, as far as we know, there is no analytical model that takes into account risk and correlation dispersion among cells. This article fills this gap. Under the Asymptotic Single Risk Factor (ASRF) assumption, we obtain new results about the bank's capital charge sensitivity to the critical parameters of the model. In particular, we show that the capital charge is not that sensitive to correlation dispersion, and the constant correlation assumption is robust.

This new result is obtained with few specifications, and we conjecture that it remains valid, at least qualitatively, for real bank portfolios that have a finite number of cells. We believe that our approach is also relevant for pioneering a new way to compute capital charges and challenge internal model assumptions as exemplified in this paper.

The plan of this paper is the following. We first provide some real data evidence about cell loss distributions and correlations. Second, we solve the ASRF model with lognormal losses at cell level, even when individual cells have various risk profiles. Third, we solve the case of non equal correlations between cells and provide some key results about the capital charge sensitivity to the main critical parameters of the model.

1. Some empirical facts about cell loss distributions and correlations

In the LDA framework, the aggregate operational loss for cell number $i$ is equal to the sum of individual losses:

$$L_i = \sum_{n=1}^{N_i} X_n^i \tag{1}$$

where $L_i$ is the aggregate loss of cell number $i$, $N_i$ is the number of events over 1 year, and $(X_n^i)_{1 \leq n \leq M}$ is the sequence of the individual loss severities for cell number $i$. The aggregate loss process is a compound Poisson process and, accordingly, the model is based on the following assumptions:

- The number of events and severities are independent
- Severities are independent and identically distributed random variables

1.1. Cell loss distribution parameters

Concerning loss distributions, there exist a lot of studies about individual loss distributions (see for instance Dutta and Perry (2004), Moscadelli (2004)), but there are very few empirical studies about aggregate cell losses.

We have conducted such a study based on the SAS OpRisk Global database. As of November 2013, this database includes 6402 events that have occurred in financial firms since 2002, date from which financial institutions started to collect and report their operational losses systematically. We have calibrated the frequency of events and lognormal severity distributions for each of the 21 cells that have more than 30 losses. Direct calibration of the aggregate loss distribution from real data is of course impossible because there is only one observation per year. However, it is possible to assess the compliance with the lognormal distribution of the aggregate loss distribution obtained through the LDA.

Let's consider that the loss distribution for cell number $i$ is lognormal with parameters $\mu_i$ and $\sigma_i$; the ratio between the expected value and any quantile depends only on the parameter $\sigma_i$:

$$\frac{\text{Expected value }(i)}{VaR_q(i)} = e^{\frac{\sigma_i^2}{2} + \sigma_i F_q} \tag{2.1}$$

$$\sigma_i = -F_q - \sqrt{F_q^2 + 2 \ln \frac{\text{Expected value }(i)}{VaR_q(i)}} \tag{2.2}$$

where $VaR_q$ is the $q$-percentile of the lognormal distribution and $F_q = N^{-1}(1-q)$. Inverting eq. (2.1) leads to two different solutions: we have chosen the one with a minus sign in front of the square root in eq. (2.2) because we require the parameters $\sigma_i$ to decrease with the ratio expected value over quantile for all cells. We observe that broader distribution assumptions for cell losses in the model can naturally be taken into account by choosing the plus sign solution in eq. (2.2).

The LDA leads to these ratios for each cell in the tail of the loss distribution ($q \geq 95\%$). For several values of the confidence level, Table 1 provides the observed average value and standard deviation of parameters $\sigma_i$ over all cells, implied from eq. (2.2).

| Confidence level | Average | StDev |
|---|---|---|
| 95% | 98% | 41% |
| 97,5% | 99% | 39% |
| 99% | 107% | 44% |
| 99,5% | 112% | 46% |
| 99,9% | 124% | 48% |
| All | 107% | 42% |

Table 1: Parameter $\sigma$ implied value from real data

The parameters $\sigma_i$ range of values is rather stable when the confidence level changes: the average value over all cells and confidence intervals is equal to 107%, and the observed standard deviation is equal to 42%. To assess the robustness of these estimates, we compute the median of observed values for the parameters $\sigma_i$ which is equal to 108.5% and is very close to the average value, and the med-med estimator (median value of the spread with the median) is equal to 31% which is lower than the measured standard deviation.

1.2. Cell loss correlations

For most of the authors (see for instance Aue and Kalkbrener, Frachot *et al.*), cell loss correlations are generated by the dependence of the number of events between cells rather than the dependence of severities. Under the assumption of lognormal severity distributions ($X^i \sim LN(m_i, s_i)$), Frachot *et al.* show that the loss correlation between cell 1 and cell 2 is equal to:

$$\text{corr}(L_1, L_2) = \text{corr}(N_1, N_2) . e^{-\frac{1}{2}s_1^2 - \frac{1}{2}s_2^2} \qquad (3)$$

The correlation of the number of events is linked to the loss frequencies of cells 1 and 2. Bivariate Poisson variables are obtained by considering three independent Poisson variables $Z, Y_1$ and $Y_2$ with parameters $r, \lambda_1 - r$ and $\lambda_2 - r$ respectively; the variables $N_i = Z + Y_i$ are also Poisson with intensities $\lambda_i$, and their correlation is equal to:

$$\text{corr}(N_1, N_2) = \frac{r}{\sqrt{\lambda_1 \lambda_2}} \leq R = \sqrt{\frac{\min(\lambda_1, \lambda_2)}{\max(\lambda_1, \lambda_2)}} \qquad (4)$$

The upper bound $R$ for the correlation comes from the inequalities $\lambda_1 \geq r$ and $\lambda_2 \geq r$. Whenever the bank's portfolio includes a large number of cells, the intensities are distributed as a random variable. Internal data better represent frequencies than external data because they are specific to the bank, and they also include frequencies of rare but severe events taken into account by scenario analyses in the model. Internal data and scenario analysis frequencies at SG support the normal distribution assumption of the log-intensities of the Poisson processes (in particular the skewness and normalized kurtosis are close to 0) with a standard deviation equal to $\gamma = 2.35 \pm 0.35$. Setting $\ln \lambda_i = \alpha + \gamma G_i$, where $(G_i)_{i=1,2}$ are uncorrelated standard normal random variables, we get:

$$R = e^{-\gamma|G_1 - G_2|/2} = e^{-\gamma|X|/\sqrt{2}}$$

where X is a standard normal random variable. Under this assumption, the upper bound $R$ follows a truncated lognormal law:

$$P[R \leq \rho] = P\left[|X| \geq -\sqrt{2}\frac{\ln \rho}{\gamma}\right] = 2 N\left(\sqrt{2}\frac{\ln \rho}{\gamma}\right) \quad (5)$$

We plot the density function of the correlation upper bound of $R$ corresponding to $\gamma = 2.35$:

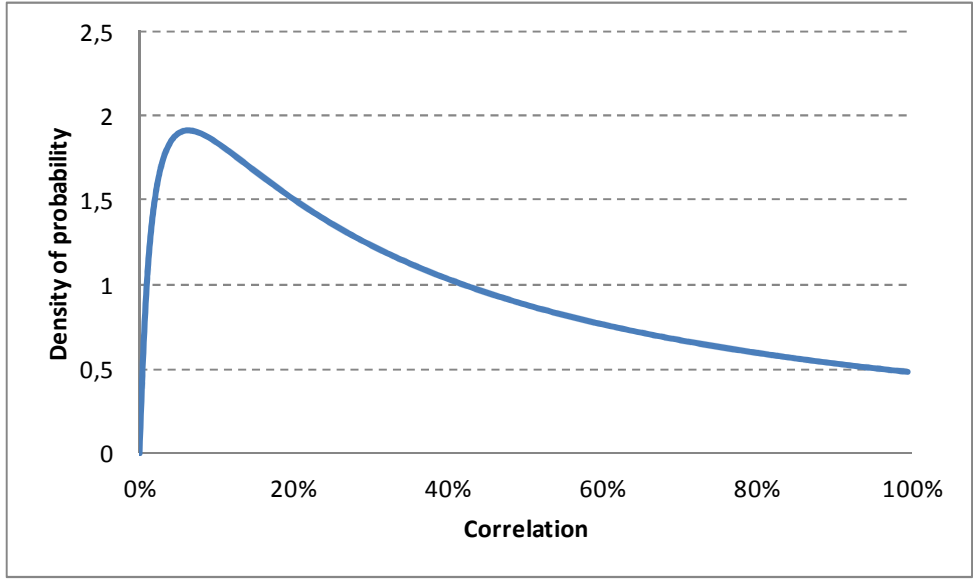

Figure 1: Density of probability of the upper bound $R$ for loss correlations

The expected value of $R$ is equal to $2e^{\sigma^2/4}N(-\gamma/\sqrt{2})$, is equal to 38.5% for $\gamma = 2.35$. This is in line with Aue and Kalkbrener (2007) who observed that frequency correlations were around 10%, and higher correlations were specific to some couples of cells only. Frachot et al. claimed that loss correlations were as low as 4%: we recover this result when we take $\text{corr}(N_1, N_2) = 38.5\%$ and $s_1 = s_2 = 1.5$ in eq. (1), which is the lowest value observed by Frachot et al. for these parameters. From SAS OpRisk data, we observe that the parameters $s_i$ have an average value equal to 2.03, a standard deviation equal to 0.42 and are ranged between 1.34 and 2.90. Correlation upper bounds can be computed with eq. (3) and (4) from these data. We find an expected value equal to 1.33% and a standard deviation equal to 1.61%. Correlation upper bounds all range between 0% and 4%, except a few of them; the highest correlation has an upper bound at 11.27% and is found between "Execution, delivery and process management" and "internal fraud" cells of the retail brokerage business line. All these studies confirm that we expect low levels of correlation between cells.

1.3. Correlation parameters in the Gaussian copula model

In the Gaussian copula framework with lognormal marginal cell losses, the correlation parameter $\rho_{ij}$ between two cells is related to the cell loss correlation:

$$\text{corr}(L_i, L_j) = \frac{e^{\rho_{ij}\sigma_i\sigma_j} - 1}{\sqrt{\left(e^{\sigma_i^2} - 1\right)\left(e^{\sigma_j^2} - 1\right)}} \quad (6)$$

This formula with parameters $\sigma_i = \sigma_j = 107\%$ and a conservative assumption for loss correlations $\text{corr}(L_i, L_j) = 4\%$, leads to a correlation parameter of the Gaussian copula equal to $\rho_{ij} = 7.2\%$. External data support the assumption of very low correlation parameters in the copula framework, much lower than 10%.

2. A class of solvable models with correlated risks

2.1. A simplified LDA model

In the rest of this article, we build a simple portfolio model for operational risk. We assess that the bank's operational risks is a portfolio of $N$ operational risks at cell level. We make the following four assumptions:
- Lognormal distributions: the loss for cell number $i$ is a lognormal random variable $L_i$ with parameters $\mu_i$ and $\sigma_i$. As shown in section 1, we assume that the $\sigma_i$ parameters have an expected value equal to $\sigma = 107\%$ and, unless otherwise stated, a variance $v = 18\%$.
- Gaussian copula: pair-wise correlations $\rho_{ij}$ may be different to each other. For numerical estimations, we assume that the average correlation is equal to 10% (this is a conservative assumption as seen in section 1.3).
- One factor model: Cell losses are sensitive to the same systemic factor called $F$. This factor is assumed to be a standard normal random variable. The specific part of the risk is embedded in another independent normal random variable called $\epsilon_i$ ($i = 1, \dots, N$). Systemic and specific factors are all assumed to be independent to each other.
- We assume that the parameters are not dependent on the number of cells $N$.

In this framework, the annual loss for a cell can be written as the exponential function of a normal random variable which is a linear combination of the systemic and specific factors. We get for cell number $i$ ($i = 1, \dots, N$):

$$L_i = e^{\mu_i - \sigma_i \left(\beta_i F + \sqrt{1-\beta_i^2}\, \epsilon_i\right)} \tag{7}$$

The parameters $\beta_i$ are linked to the pair-wise correlations of the Gaussian copula: $\rho_{ij} = \beta_i . \beta_j$. Because cells may have very different risk characteristics, and because correlations may be very different for different pairs of cells, we assume that the parameters $\mu_i, \sigma_i, \beta_i$ are the observations of *i.i.d.* random variables called $M, \Sigma, B$ respectively. In the limit $N \to \infty$, the bank's loss is equal to $N.L(F)$ and is a function of the common factor $F$, as in Vasicek's model for granular homogeneous loan loss distributions (see Vasicek, 2002):

$$L(F) = \lim_{N \to \infty} \frac{1}{N} \sum_{i=1}^{N} L_i = E\left[e^{M - \Sigma\left(BF + \sqrt{1-B^2}\epsilon_i\right)} \middle| F\right] = E\left[e^{-\Sigma BF + \Sigma^2 (1-B^2)/2} \middle| F\right].E[e^M] \tag{8}$$

Without loss of generality, we assume $M = 0$ because it simply rescales the bank's loss by a constant factor $E[e^M]$ in the $N \to \infty$ limit.

The stand-alone capital for cell number $i$, called $KSA_i$, is equal to the 99.9% percentile of the cell loss distribution and is equal to $KSA_i = e^{-\sigma_i F_q}$, where $F_q = N^{-1}(0.1\%)$. The bank's capital charge is equal to $N.L(F_q)$. The capital reduction coming from risk diversification is measured by the Diversification Index defined as:

$$\text{Diversification Index} = DI = \frac{N.L(F_q)}{\sum_{i=1}^{N} KSA_i} \xrightarrow{N \to \infty} \frac{L(F_q)}{E[e^{-\Sigma F_q}]} \tag{9}$$

## 2.2. Homogeneous risks

The simplest solvable model is obtained for homogeneous risks: The random variables $\Sigma$ and $B$ have a constant value equal to $\sigma$ and $\sqrt{\rho}$ respectively for all cells, and $v = 0$. In the limit $N \to \infty$, the bank's loss distribution remains lognormal:

$$L(F) = \lim_{N \to \infty} \frac{1}{N} \sum_{i=1}^{N} L_i = e^{-\sigma\sqrt{\rho}F + \sigma^2(1-\rho)/2} \tag{10}$$

As in Markowitz's portfolio theory, risk is not diversified away because of loss correlations between cells. The correlation parameter determines the Diversification Index (*DI*):

$$DI = e^{\sigma(1-\sqrt{\rho})F_q + \sigma^2(1-\rho)/2}$$

For $\sigma = 107\%$ and $\rho = 10\%$, we get $DI = 17.5\%$. We note that $DI > 1$ when $\sigma > -2F_q \frac{1-\sqrt{\rho}}{1-\rho} = 4.70$ and capital charges are no longer sub-additive. However, super-additivity occurs for values of the parameter $\sigma$ higher than the typical value of 107%. As explained in section 1, broader distributions can easily be accommodated in our model by choosing the other solution of eq. (2.2). This would similar effects as with fait tail distributions (Neslehova *et al.*, 2006).

The diversification ratio is particularly low because of the $N \to \infty$ limit. The perimeter and number of cells is however a modeling choice and a convention. Choosing a very high number of cells would not necessarily result in a regulatory arbitrage; to assess this, the scaling of the parameters of the model with the number of cells must be investigated.

## 2.3 Heterogeneous risk and identical correlations

In reality, cells have different risk characteristics. We assume that the parameters $\sigma_i$ are normally distributed, i.e. $\Sigma \equiv N(\sigma, v)$, and that correlations are constant, i.e. $B = \sqrt{\rho}$. The loss at cell level writes:

$$L_i = e^{-\sigma_i(\sqrt{\rho}F + \sqrt{1-\rho}\epsilon_i)}$$

In the limit $N \to \infty$, the bank's loss writes from (8) and (11) as a Gaussian integral:

$$L(F) = \int_{-\infty}^{\infty} \frac{dx}{\sqrt{v}} n\left(\frac{x-\sigma}{\sqrt{v}}\right) e^{-x\sqrt{\rho}F + (1-\rho)x^2/2} = \frac{1}{\sqrt{1-(1-\rho)v}} e^{-\sigma\sqrt{\rho}F + \sigma^2(1-\rho)/2 + \frac{v}{2}\frac{\left((1-\rho)\sigma - \sqrt{\rho}F\right)^2}{1-(1-\rho)v}} \tag{11}$$

where $n(x) = e^{-x^2/2}/\sqrt{2\pi}$. The bank's loss follows a g-and-h distribution. Because we assumed that the random variable $\Sigma$ is normal and, strictly speaking, could take negative values, the resulting bank's loss is a non decreasing function of the systemic factor. However, this occurs when the systemic factor $F$ is larger than $F^* = \sigma/v\sqrt{\rho}$ which is very unlikely in practice (for instance, $F^* = 18.8$ when $\rho = 10\%$, $\sigma = 107\%$ and $v = 18\%$); the normal law assumption for $\Sigma$ is thus suitable, in particular to model the tail of the loss distribution, whenever $F \ll 0$. We show in section 3 that the shape of the parameters' distribution function is not critical.

The bank's capital charge, $N.L(F_q)$, increases with $v$. For $\rho = 10\%$, $\sigma = 107\%$ and $v = 18\%$, this increase is equal to +62%. Unsurprisingly, the capital charge is very sensitive to risk dispersion measured by the parameter $v$, which is a critical parameter of the model. Changing the value of $\rho$ from 10% to 20% leads to a capital charge increase of +62%. The average correlation level is a critical parameter of the model as well. We notice that, as the sum of the stand-alone capital charges is equal to $N.E[e^{-\Sigma F_q}] = N.e^{-\sigma F_q + vF_q^2/2}$, the resulting $DI$ is a decreasing function of $v$. Non equal parameters $\sigma_i$, when included as an uncorrelated additional risk, increase the capital charge but generate more diversification, as illustrated in fig. 2.

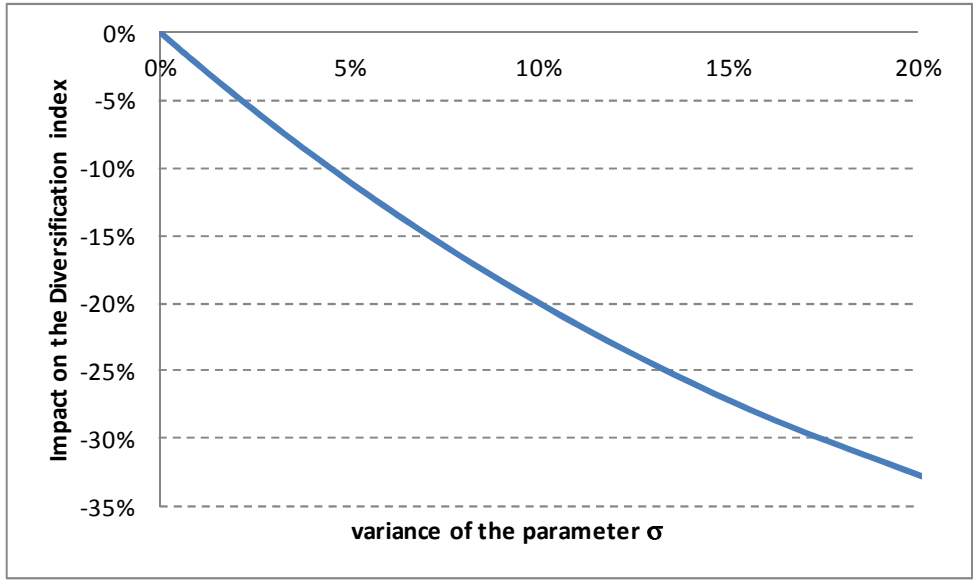

Figure 2: Diversification index impact as a function of $\sqrt{v}$ ($\sigma = 107\%, \rho = 10\%$).

3. Uncertain correlations

Correlations are not identical to each other, as illustrated in section 1, but estimating them from real data is a challenge from a statistical viewpoint. Data are scarce and limited only to one observation per year for the aggregate loss. Estimation of the number of events correlation is no longer robust for the same reason and severity correlations are only observable for cells that exhibit a sufficient number of events per year. Assuming identical correlations among cells is a current practice even if, in reality, correlations are unknown parameters. In what follows, we remain in the limit $N \to \infty$ and correlation uncertainty is included in the model by assuming that the random variable $B$ has an expected value equal to $\beta = \sqrt{\rho}$ and a variance equal to $w$. For the sake of clarity we assume that individual risks are all equal among cells, *i.e.* $\Sigma$ is a constant equal to $\sigma$. We obtain, in the limit $N \to \infty$, from (8) and (15):

$$L(F) = \lim_{N \to \infty} \frac{1}{N} \sum_{i=1}^{N} L_i = E\left[e^{-\sigma BF + \sigma^2(1-B^2)/2} | F\right] = \int_{-\infty}^{\infty} dx\, f(x) e^{-x\sigma F + (1-x^2)\sigma^2/2} \qquad 15)$$

Where the function $f(.)$ is the density of the random variable $B$. If we assume that the variable $B$ is normally distributed ($B \equiv N(\beta, w)$), we get:

$$L(F) = \frac{1}{\sqrt{1+\sigma^2 w}} e^{-\beta \sigma F + (1-\beta^2)\sigma^2/2 + \frac{\sigma^2 w}{1+\sigma^2 w}(\beta \sigma + F)^2/2} \qquad (16)$$

If the variable $B$ is uniformly distributed between $\beta - \sqrt{3w}$ and $\beta + \sqrt{3w}$ (the bounds are chosen so that the expected value and the variance are equal to $\beta$ and $w$ respectively), we get:

$$L(F) = \sqrt{\frac{\pi}{6w\sigma^2}} e^{\sigma^2/2 + F^2/2} \left[N\left(\sigma(\beta + \sqrt{3w}) + F\right) - N\left(\sigma(\beta - \sqrt{3w}) + F\right)\right] \qquad (17)$$

As pair-wise correlations are equal to $\rho_{ij} = \beta_i.\beta_j$, there is a direct link between the variance of $\rho_{ij}$ and the variance of the sensitivity parameters $\beta_i$. Because of the independence of the $\beta_i$, we have:

$$var(\rho_{ij}) = E[\beta_i^2.\beta_j^2] - E[\beta_i]^2.E[\beta_j]^2 = w(w + 2\beta^2)$$

This leads, by solving the second order equation in $w$, to:

$$w = \sqrt{\beta^4 + var(\rho_{ij})} - \beta^2 \qquad (18)$$

For $\beta^2 = \rho = 10\%$ and $\sqrt{var(\rho_{ij})} = 3\%$ (which is a conservative value compared to what is measured from observed data; see section 1.2), we have $w = 0.44\%$, i.e. the standard deviation of the parameter $\beta$ is equal to 6.6%. The ratio of the capital charge including model risk ($w > 0$) and the capital charge without model risk ($w = 0$) measures the increase in capital due to dispersion or uncertainty on correlations. We plot this quantity in figure 3 as a function of the mean-deviation of the correlation parameter ($\sqrt{w}$); we show that the impact of the mean-deviation of the correlation parameter is lower than 2% for $\sqrt{w} = 6.6\%$ for both the normal and uniform assumptions. Additionally, as the curves are very close to each to other, we conclude that the shape of the correlation distribution function is not a driver of the capital charge: this validates the choice of the normal law for distribution functions that we have done throughout this article.

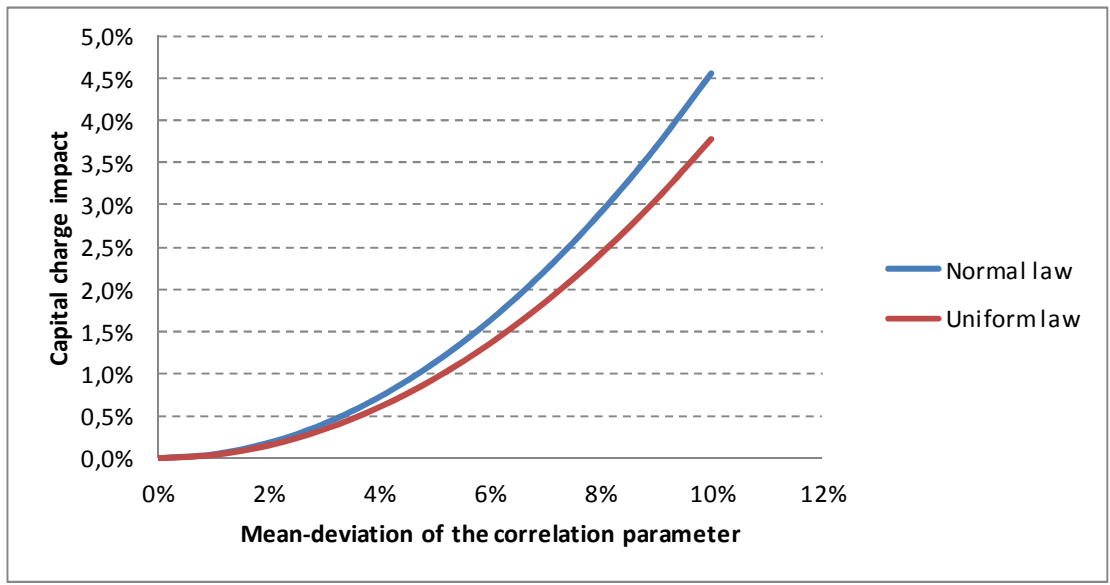

Figure 3: Capital charge impact as a function of the correlation dispersion for normal and uniform laws ($\beta^2 = 10\%$ and $\sigma = 107\%$)

Even with a much more conservative choice for the individual cell risk parameter $\sigma = 200\%$, the impact of correlation dispersion on the bank's capital charge would be around +5%. Our conclusion is that correlation dispersion (measured by parameter $w$) is, by far, not as critical as the other parameters of operational risk models (average cell risk $\sigma$, cell risk dispersion $v$ and average correlation level $\beta = \sqrt{\rho}$).

Discussion and conclusion

This article pioneers analytical models for computing bank's operational risk capital charges and provides some new results on the correlation problem. These simplified models are somehow realistic as they incorporate dispersion in individual cell risks and correlation levels. The main result of this paper is that uniform correlation is a robust assumption for capital charge modeling. This result is important because it means that model risk associated with the value of correlations is not a major issue for capital measurement. The impacts of the choice of the copula function and of the average correlation value are much more significant, albeit calibration suffers from the scarcity of observed data. At the end of the day, dependence appears as a subjective choice that determines the diversification benefit at the bank's level, and cell loss distribution functions remain the main driver of the capital charge. We emphasize that our approach can straightforwardly be extended to other cell loss distribution or copula functions (Student for instance), in the one factor framework. The extensions of our analytical approach should, as a second step, focus on broad distribution functions for cell losses and smaller number of cells. This is left for future research.

The author wants to thanks the two anonymous referees for their comments on the paper. Special thanks to Pavel Shevchenko as well for the very stimulating correspondence we had together. This article reflects the author's opinions and not necessarily those of his employers.

References

Aue, F. and M. Kalkbrener, LDA at work: Deutsche Bank's Approach to Quantifying Operational Risk, Journal of Operational Risk, Vol. 1 (2007) 49–93

Bank for International Settlements, Operational Risk – Supervisory Guidelines for the Advanced Measurement Approaches, BCBS 196 (2011)

Chernobai, A., Rachev, S.T. and Fabozzi, F., Operational Risk: A Guide to Basel II Capital Requirements, Models and Analysis, John Wiley & Sons (2007)

Dutta, K. and Perry, J., A Tale of Tails: An Empirical Analysis of Loss Distribution Models for Estimating Operational Risk Capital, The Federal Reserve Bank of Boston, working paper (2007)

Frachot, A., T. Roncalli and E. Salomon, The Correlation Problem in Operational Risk, Groupe de Recherche opérationnelle, Crédit Agricole, France (2004)

Moscadelli, M., The modelling of operational risk: experience with the analysis of the data collected by the Basel committee. Technical Report 517, Banca d'Italia (2004)

Neslehova, J., Embrechts, P., Chavez-Demoulin, V.: Infinite Mean Models and the LDA for Operational Risk, Journal of Operational Risk **1,** 1 (2006) 3-25

Vasicek O, Loan Portfolio Value, Risk December (2002) 160–162